\documentstyle[11pt]{article}
\begin{document}
\title{Miura Map between Lattice KP and its Modification is Canonical}

\author{Q.P. Liu\\
The Abdus Salam International Centre for Theoretical Physics\\
Trieste 34100, Italy\\
and \\
Beijing Graduate School\\
China University of Mining and Technology\\
100083, Beijing\\
P.R. China\footnote{Mailing address}}
\date{}
\maketitle{}
\begin{abstract}
We consider the Miura map between the lattice KP hierarchy and the lattice modified
KP hierarchy and  prove that the map is canonical not only between the
first Hamiltonian
structures, but also between the second Hamiltonian structures. 
\end{abstract}

\section{Introduction}
It is well-known that Miura map, a transformation between the KdV equation and
MKdV equation, plays a central role in the development of soliton theory. Indeed,
the celebrated Inverse Scattering Method for solving nonlinear equations starts
with the Miura map \cite{miura}. This type of transformations turns out to exist in the
context of other integrable equations (see \cite{af}-\cite{liu1}\cite{shaw}-\cite{st} and the references there).

Kupershmidt, in a recent paper \cite{bak2}, considered the canonical properties of Miura
maps
between KP and MKP hierarchies. He shown that, both in continuous and discrete cases,
 Miura transformations are canonical between the first Hamiltonian structures.
For the ordinary or continuous KP and its modification, Shaw and Tu \cite{shaw} generalized
the results
of Kupershimdt and
proved that the very Miura map is also canonical between the second Hamiltonian
structures.

We will consider the canonical property of the Miura map between the Lattice MKP
(lMKP) and the Lattice KP (lKP) hierarchies. The lKP hierarchy is a bi-Hamiltonian
system and two Hamiltonian structures were constructed by using the residue calculus
in \cite{bak1}. For the lMKP hierarchy, the first Hamiltonian structure was also
found in
\cite{bak2}. A slight different version of the lMKP hierarchy was proposed by Oevel
and he
further obtained the bi-Hamiltonian description for this hierarchy by means of
r-matrix approach \cite{oevel}. By introducing a parameter, we unify Kupershmidt's version
of the lMKP hierarchy
and Oevel's version into a single system. Our main
purpose of the paper is to prove that Kupershmidt's Miura map
is a canonical map not only between the first Hamiltonian structures of lKP and lMKP, but
also
between the second Hamiltonian structures.

The paper is organized as follows. In the next section, we introduce notations
and recall the relevant formulae such as bi-Hamiltonian structures of the lKP
and lMKP hierarchies.  In section 3 and section 4, we show that Kupershmidt's Miura map is 
a canonical transformation for the first Hamiltonian structures and the second 
Hamiltonian structures respectively. The last section is intended to summary and discussions.

\section{Background and Notations}
To introduce the lKP and lMKP hierarchies, we consider the algebra of shift operators
$$
g=\big\{u_N(n)T^N+u_{N-1}(n)T^{N-1}+\cdots+u_0(n)+u_{-1}(n)T^{-1}+\cdots\big\},
$$
where $u_j$ are scalar functions of integer $n$. The shift operator $T$ is
given by
$$
(Tf)(n)=f^{(1)}(n):=f(n+1),
$$
and for arbitrary integer $k$, $(T^kf)(n)=f^{(k)}(n)=f(n+k)$.

For any operator $\xi=\sum_{j}u_jT^j\in g$, the projections to various shift orders are
denoted by
$$
\xi_j=u_jT^j, \;\;\;\; \xi_{\geq k}=\sum_{j\geq k}u_jT^j,\;\;\;\;
\xi_{< k}=\sum_{j<k}u_jT^j,
$$
$$
\xi_{>k}=\sum_{j>k}u_jT^j,\;\;\;\;\;\; \xi_{\leq k}=\sum_{j\leq k}u_jT^j.
$$
From the shift operator $T$, we also have the difference operator
$$
\Delta=T-1,
$$
and its formal inverse
$$
\Delta^{-1}=\sum_{j\geq 1}T^{-j}.
$$
Another important notation is so called the trace, which is defined as
$$
{\rm{tr}}(\sum_{i}u_iT^i)=\sum_{n} u_0(n),
$$
this permits us to
identify $g$ and its dual by the metric
$g^*$: $<u^*,u>={\rm{tr}}(u^*u)$. It can be shown that the metric is
bi-invariant.

The lKP hierarchy is defined by the following Lax operator 
\begin{equation}
L=T +\sum_{i=0}^{\infty}A_i T^{-i},
\end{equation}
and the flow equations are constructed as
\begin{equation}
L_{t_n}=[(L^{n})_{\ge 0},L].
\label{lkp}
\end{equation}
The lKP hierarchy (\ref{lkp}) is a bi-Hamiltonian system. Its two
Hamiltonian structures are
constructed by means of the residue calculus in \cite{bak1}. Recently,
Oevel proposed
a $r-$matrix setting for these Hamiltonian structures. The two Hamiltonian structures
are given by the following Poisson 
tensors
\begin{eqnarray}
P_1(\nabla H)&=&[\nabla H, L]_{\leq 0},\label{tensor11}\\
P_2(\nabla H)&=&(L\nabla H)_{\geq 1}L-L(\nabla HL)_{\geq 1}+\nonumber\\
             & & \frac{1}{2} [(L\nabla H+\nabla HL)_0,L]+
              \frac{1}{2}[\rho([\nabla H,L]_0),L],
\label{tensor12}
\end{eqnarray}
where $\rho$ is a skew-symmetric linear map on the algebra $g_0$ given explicitly 
 by
\begin{equation}
\rho=\frac{T+1}{T-1},
\label{ro}
\end{equation}
and
\begin{equation}
\nabla H=\frac{\delta H}{\delta A_0} +T\frac{\delta H}{\delta A_1}+\cdots .
\label{h}
\end{equation}
As for the lMKP hierarchy, we consider the following Lax
operator
\begin{equation}
{\cal L}=qT+\sum_{i=0}^{\infty}a_i T^{-i},
\end{equation}
and the flow equations are represented by
\begin{equation}
{\cal L}_{t_{n}}=[({\cal L}^n)_{\geq 1}+\alpha({\cal L}^n\Delta^{-1})_0, {\cal L}],
\label{lmkp}
\end{equation}
where $\alpha$ is a constant.

The case $\alpha=0$ and the case $\alpha=1$ were considered by Kupershmidt and Oevel
respectively. In these two cases, the mlKP hierarchy (\ref{lmkp}) is a
bi-Hamiltonian system. When
$\alpha=0$, the
first Haimltonian structure of (\ref{lmkp}) is found by
Kupershimdt in the context of the residue calculus, it is not clear how to construct the
second
one this way. Oevel, in the case $\alpha =1$, gives the bi-Hamiltonian structures 
by means of $r-$matrix approach.

Consider the linear operator on $g$
\[
r(\xi)=\xi_{\geq 1}-\xi_{<1}-2\alpha(\xi\Delta^{-1})_0,
\]
by direct calculations, it is found that that  above $r$ solves the modified
Yang-Baxter equation only and only if $\alpha=0$ or $\alpha=1$. As we mentioned above, these
are exactly the
two cases studied by  Kupershimdt and by Oevel. In the following, our parameter  $\alpha$ 
will take the value either one
or zero. The above $r$-matrix leads to the first Hamiltonian structure for
the lMKP hierarchy.

To get the second Poisson tensor, one may use Suris's construction \cite{suris} by
considering the following linear operators
\begin{eqnarray*}
{\cal A}_1(\xi)&=&\xi_{\geq 1}-\xi_{<0}-2\alpha (\xi\Delta^{-1})_0-\rho(\xi_0)
       +2\alpha\Delta^{-1}\xi_0,\\
{\cal A}_2(\xi)&=&\xi_{\geq 1}-\xi_{<0}+\rho(\xi_0),\\
{\cal S}(\xi)&=&\rho(\xi_0)-\xi_0-2\alpha \Delta^{-1}\xi_0,\\
{\cal S}^{\dagger}(\xi)&=&-\rho(\xi_0)-\xi_0-2\alpha (\xi \Delta^{-1})_0.
\end{eqnarray*}
When $\alpha=1$, the above operators are those presented by Oevel and are lead
to the second Poisson tensor for this case. It can be proved that in the case
$\alpha=0$, these operators are satisfied the conditions of Suris's theorem
(or the theorem 1 of
Oevel \cite{oevel}),
therefore they  also lead to a Poisson tensor, this time is for Kupershimidt' case.
Unifying both Kupershmidt's case and Oevel's case, we have the following
two Poisson tensors 
\begin{eqnarray}
{\tilde{\cal P}}_1(\nabla H)&=&[(\nabla H)_{\geq 1}, {\cal L}]-
             [\nabla H, {\cal L}]_{\geq 0}-
             \alpha[(\nabla H\Delta^{-1})_0,{\cal L}]-
             \alpha\Delta^{-1}[\nabla H,{\cal
L}]_0,\label{tensor21}\\
{\tilde{\cal P}}_2(\nabla H)&=&({\cal L}\nabla H)_{\geq 1}{\cal L}-{\cal L}(\nabla H{\cal L})_{\geq 1}+
            +\frac{1}{2}[{\cal L}, \nabla H]_0{\cal L}+\frac{1}{2}{\cal
L}[{\cal L},\nabla H]_0+ \nonumber\\
          & &  \alpha\Delta^{-1}[{\cal L},\nabla H]_0{\cal L}+
            \alpha[{\cal L}, ({\cal L}\nabla H\Delta^{-1})_0]+
            \frac{1}{2}[\rho([\nabla H, {\cal L}]_0),{\cal L}],
\label{tensor22}
\end{eqnarray}
where $\rho$ is the one defined by (\ref{ro}) and $\nabla H$ is
parametrized as
\begin{equation}
\nabla H=T^{-1}\frac{\delta H}{\delta q}+\frac{\delta H}{\delta a_0}+
         T\frac{\delta H}{\delta a_1}+\cdots
\label{h1}
\end{equation}

In the remaining part of this section, we introduce the Miura map between lKP hierarchy and
lMKP hierarchy following Kupershmidt. With the aid of a new field
$w$,  we introduce a 
map between lKP and mlKP hierarchies via the conjugacy
$$
L={\rm e}^w{\cal L}{\rm e}^{-w}={\rm e}^wqT{\rm e}^{-w}+
\sum_{i=0}^{\infty}{\rm e}^w a_i T^{-i}{\rm e}^{-w},
$$
comparing the coefficients of different power of the shift operator of two sides leads to
a transformation
$$
q={\rm e}^{w^{(1)}-w},\;\;\;\; A_i=a_i{\rm e}^{w-w^{(-i)}}, \;\; (i\geq
0).
$$
Let us introduce the new notations
$$
R_i=R_i(q):=\prod_{s=0}^{i}q^{(-s)}/q, \;\; (i\geq 0).
$$
By eliminating the intermediate variable $w$, we reach  the Miura map
between the two sets of
variables
\begin{equation}
M:\;\;\; A_0=a_0, \;\;\;\; A_i=R_ia_i \;\; (i>0).
\label{miura}
\end{equation}
this is the Miura map constructed in \cite{bak2}.

Now we prove that if ${\cal L}$ solves the mlKP hierarchy,
$L={\rm e}^w{\cal L}{\rm e}^{-w}$
solves the lKP hierarchy.
From $q={\rm e}^{w^{(1)}-w}$, we obtain $q_t=q(T-1)w_t$. On the other hand,
the time evolution of $q$ can be read from the mlKP hierarchy, that is
$ q_t=q(T-1)\Big(({\cal L}^n)_0+\alpha ({\cal L}^n\Delta^{-1})_0\Big)$, so
$w_t=({\cal L}^n)_0+\alpha ({\cal L}^n\Delta^{-1})_0$. Now
\begin{eqnarray*}
L_t&=&[w_t,L]+{\rm e}^w[({\cal L}^n)_{\geq 1}-
    \alpha ({\cal L}^n\Delta^{-1})_0,{\cal L}]{\rm e}^{-w}\nonumber\\
   &=&[w_t,L]+[{\rm e}^w({\cal L}^n)_{\geq 1}{\rm e}^{-w},L]-
      \alpha [({\cal L}^n\Delta^{-1})_0, L]
    =[(L^n)_{\geq 0},L],
\end{eqnarray*}
where we used $ ({\cal L}^n)_0=(L^n)_0$ and $e^w({\cal L}^n)_{\geq 1}=
(L^n)_{\geq 1} e^w$. Thus the Miura map (\ref{miura}) indeed converts the
lMKP hierarchy
into the lKP hierarchy.

\section{Canonical Properties for First Hamiltonian Structures}
In this section, we prove that Miura map is canonical between the
first Hamiltonian structures. First we calculate the Hamiltonian matrices from
the Poisson tensors (\ref{tensor11}) and (\ref{tensor21}). By substituting
(\ref{h}) into $P_1$ and (\ref{h1}) into $P_2$, it is
straightforward to get 
\begin{equation}
B_{1}^{lKP}=(B_{ij}), B_{ij}=T^jA_{i+j}-A_{i+j}T^{-i}, \;\; (i,j\geq 0),
\end{equation}
and 
\begin{equation}
B_{1}^{lMKP} =\bordermatrix{ &q& a_0 & a_{j>0}\vspace{0.3cm}\cr q&0
&q(T-1) &\alpha q(T-1)T^j\cr
a_0&(1-T^{-1})q&0&0\cr a_{i>0}&\alpha T^{-i}(1-T^{-1})q&0&B_{ij}^{(lMKP)}},                    
\end{equation}
where
\begin{equation}
B_{ij}^{(lMKP)}=T^ja_{i+j}-a_{i+j}T^{-i}+
\alpha(a_iT^{j-i}-T^{j-i}a_j+T^{-i}a_j-a_iT^{j}).
\end{equation}
The Jacobian matrix of the Miura map (\ref{miura}) is easily calculated as
\begin{equation}
J=\bordermatrix{ &q&a_0&a_{i>0}\vspace{0.2cm}\cr A_n
&
a_nD_n&R_n\delta_{n}^{0}&R_n\delta_{n}^{i} },
\end{equation}
where $\delta_{j}^{i}$ is the standard Kronecker symbol and $D_n$ is
the abbreviated 
notation for the Fr\'{e}chet derivative given by
\begin{equation}
D_n:=D(R_n)=R_n\frac{1-T^{-n}}{T-1}q^{-1},\;\;\;\;
D_n^{\dagger}=q^{-1}\frac{1-T^{n}}{T^{-1}-1}R_n.
\label{der}
\end{equation}

We need to calculate the matrix operator $JB_{1}^{lMKP}J^{\dagger}$, but first it is easy to
find that
\[
JB_{1}^{lMKP}=\bordermatrix{ &q&a_0&a_{j>0}\vspace{0.3cm}\cr
          A_0&(1-T^{-1})q&0&0\cr
          A_{i>0}&\alpha R_iT^{-i}(1-T^{-1})q&
          a_iD_iq(T-1)&\alpha a_iD_iq(T-1)T^j+R_iB_{ij}^{(lMKP)}},
\]
now the entries of the first row of the $JB_{1}^{lMKP}J^{\dagger}$ are seen as 
$$
(JB_{1}^{lMKP}J^{\dagger})_{0,m}=(1-T^{-1})qD_{m}^{\dagger}a_m=-(1-T^m)R_ma_m=
(T^m-1)A_m,
$$
which coincide
with the $(B_{1}^{lKP})_{0,m}$. It is noticed that we have used the
second formula of (\ref{der}). Therefore,
for the first row and the first column, two
matrix operators $B_{1}^{lKP}$ and $JB_{1}^{lMKP}J^{\dagger}$ are just
the same as expected. We
turn our attention to other entries of matrices.
We find that
\begin{eqnarray*}
(JB_{1}^{lMKP}J^{\dagger})_{mn}&=& \alpha
R_mT^{-m}(1-T^{-1})qD_{n}^{\dagger}a_n+\alpha
a_mD_mq(T-1)T^{n}R_n+\\
                    & & \alpha R_m(a_mT^{n-m}-T^{n-m}a_n+T^{-m}a_n-a_mT^n)R_n+\\
                    & & R_m(T^n a_{m+n}-a_{n+m}T^{-m})R_n                       \\
                    &=&\alpha R_mT^{-m}(T^n-1)R_na_n+\alpha a_mR_m(1-T^{-m})T^nR_n+\\
                    & &\alpha R_m(a_mT^{n-m}-T^{n-m}a_n+T^{-m}a_n-a_mT^n)R_n+\\
                    & & R_m(T^na_{m+n}-a_{n+m}T^{-m})R_n\\
                    &=&R_m(T^na_{m+n}-a_{n+m}T^{-m})R_n,
\end{eqnarray*}
now we use the formula in \cite{bak2} 
$$
R_nT^mR_m=T_mR_{n+m},
$$
and obtain the desired the results
$(JB_{1}^{lMKP}J^{\dagger})_{mn}=(B_{1}^{lKP})_{mn} $.
Thus, Miura map is indeed canonical.

\section{Canonical Property for Second Hamiltonian Structures}
We now show that the Miura map (\ref{miura}) is also canonical between the
second Hamiltonian structure
of the lKP hierarchy and the second Hamiltonian structure of the lMKP hierarchy. As in last
section, we first calculate the Hamiltonian matrix operators from the Poisson tensors
(\ref{tensor12})
and (\ref{tensor22}). The
calculation in the present case is a bit cumbersome although it is straightforward. For the lKP
hierarchy we have

\[
A_{k,t}=\sum_{\ell=0}^{\infty}(B_{2}^{lKP})_{k\ell})\frac{\delta H}{\delta A_{\ell}}, \;\;\;
k\geq 0, \;\;\; t\equiv t_n, \;\;\; H\equiv H_{n}=\frac{1}{n}{\rm tr}(L^n),
\]
where
\begin{eqnarray*}
(B_{2}^{lKP})_{k\ell}&=&\sum_{j=1}^{\ell+1}(A_{\ell-j}T^{j}A_{k+j}-A_{k+j}T^{\ell-k-j}A_{\ell-j})+\\
& &A_k(1-T^{-k})(1+T+\cdots+T^{\ell})A_l ,\;\; A_{-1}\equiv 1, \;\; k\geq 0, \ell\geq 0.
\end{eqnarray*}

For the mlKP hierarchy, we have
\[
B_{2}^{lMKP}=\bordermatrix{&q&a_{m\geq0}\vspace{0.3cm}\cr
     q&q(T-T^{-1})q&\begin{array}{r}\alpha
q(T-1)\sum\limits_{i=1}^{m+1}a_{m-i}T^i+\\
q(T-T^m)a_m\end{array}\vspace{0.2cm}\cr
  a_{k\geq 0}&\begin{array}{r}\alpha \sum\limits_{j=1}^{k+1}T^{-j}a_{k-j}(1-T^{-1})q+\\
  a_k(T^{-k}-T^{-1})q\end{array}&B_{km}^{lMKP},
 }
\]
with 
\begin{eqnarray*}
B^{lMKP}_{km}&=&\sum_{i=1}^{m+1}(a_{m-i}T^i a_{k+i}-a_{k+i}T^{m-k-i}a_{m-i})+
           a_k\frac{(1-T^{-k+1})(1-T^m)}{1-T}a_m+\\
      & & \alpha a_k(T^{-k}-1)\sum_{i=1}^{m+1}a_{m-i}T^i
+\alpha\sum_{j=1}^{k+1}T^{-j}a_{k-j}(1-T^m)a_m,
      \;\;\; a_{-1}\equiv q.
\end{eqnarray*}        
Thus, the matrix operator $JB_{2}^{lMKP}$ reads as
\[
JB_{2}^{lMKP}=\bordermatrix{&q&a_{m\geq 0}\vspace{0.3cm}\cr
        A_0&(a_0+\alpha
T^{-1}q)(1-T^{-1})q&\begin{array}{r}qT^{m+1}a_{m+1}-a_{m+1}T^{-1}q+\\
\alpha T^{-1}q(1-T^m)a_m\end{array}
\vspace{0.3cm}\cr
        A_{k\geq 0}&\begin{array}{r}a_kD_kq(T-T^{-1})q+\\ a_kR_k(T^{-k}-T^{-1})q+\\ \alpha
R_k\sum\limits_{j=1}^{k+1}T^{-j}a_{k-j}(1-T^{-1})q\end{array}
       &\begin{array}{r}a_kD_k\Big(q(T-T^{-m})a_m+\\ \alpha R_k
q(T-1)\sum\limits_{i=1}^{m+1}a_{m-i}T^i\Big)+R_k B_{km}^{lMKP}\end{array}.
}
\]
With all these formulae in hand, we find that the entries of  the first row of
$JB_{2}^{lMKP}J^{\dagger}$
are 
\begin{eqnarray*}
(JB_{2}^{lMKP}J^{\dagger})_{0,n}&=&(a_0+\alpha T^{-1}q)(1-T^{-1})qD_{n}^{\dagger}a_n+\\
                     & &\Big(qT^{n+1}a_{n+1}-a_{n+1}T^{-1}q+
                        \alpha T^{-1}q(1-T^n)a_n\Big)R_n\\
  &=&-(a_0+\alpha T^{-1}q)(1-T^n)A_n+qT^{n+1}a_{n+1}R_n-\\
  & &a_{n+1}T^{-1}qR_n+\alpha T^{-1}q(1-T_n)A_n\\
  &=&-a_0(1-T^n)A_n+qT^{n+1}a_{n+1}R_n-a_{n+1}T^{-1}qR_n=\\
  &=&A_0(T^n-1)A_n+T^{n+1}A_{n+1}-A_{n+1}T^{-1}=(B_{2}^{lKP})_{0,n},
\end{eqnarray*}
where we used 
$$
 q^{(-n-1)}R_n=R_{n+1},\;\;\;\; q^{-1}R_{n}^{(-1)}=R_{n+1},
$$
 which hold identically.
For the remaining entries, we have,
\begin{eqnarray}
\lefteqn{(JB_{2}^{lMKP}J^{\dagger})_{mn}=}\nonumber\\  
& &a_mD_mq(T-T^{-1})qD_{n}^{\dagger}a_n+R_ma_m(T^{-m}-T^{-1})qD_{n}^{\dagger}a_n+\label{j1}\\
                    & &a_mD_m q(T-T^n)a_n+R_ma_m\frac{(1-T^{-m+1})(1-T^n)}{1-T}a_nR_n+\label{j2}\\
                    & &\alpha R_m\sum_{j=1}^{m+1}T^{-j}a_{m-j}\Big((1-T^{-1})qD_{n}^{\dagger}a_n+
                       (1-T^n)a_nR_n\Big)+\label{j3}\\
                    & &\alpha \Big(a_m D_mq(T-1)+R_ma_m(T^{-m}-1)\Big)\sum_{j=1}^{n+1}a_{n-j}T^jR_n+\label{j4}\\
                    &
&R_m\Big(\sum_{l=1}^{n+1}(a_{n-l}T^la_{m+l}-a_{m+l}T^{n-m-l}a_{n-l})\Big)R_n,
\label{jb5}
\end{eqnarray}
so  we need to prove that above expression is $(B_{2}^{lKP})_{mn}$.

It is easy to see that $(\ref{j3})=(\ref{j4})=0$ in terms of $D_m$.

Since $T-T^{-1}=-(1+T)(T^{-1}-1)$,  we obtain
\begin{eqnarray*}
(\ref{j1})+(\ref{j2})&=&-A_m\frac{1-T^{-m}}{T-1}(1+T)(1-T^n)A_n+A_m(T^{-m}-T^{-1})\frac{1-T^n}{T^{-1}-1}A_n \\
    & &  +A_m\frac{1-T^{-m}}{T-1}(T-T^n)A_n +A_m\frac{(1-T^{-m+1})(1-T^n)}{1-T}A_n \\
    &=&A_m\frac{1-T^{-m}-T^{n+1}+T^{n-m+1}}{1-T}A_n=A_m\frac{(1-T^{-m})(1-T^{n+1})}{1-T}A_n.
\end{eqnarray*}
Thus to complete the proof, we need to show that
\[
R_m\big(\sum_{i=1}^{n+1}a_{n-i}T^ia_{m+i}-a_{m+i}T^{n-m-i}a_{n-i}\big)R_n=
\sum_{j=1}^{n+1}(A_{n-j}T^jA_{m+j}-A_{m+j}T^{n-m-j}A_{n-j}),
\]
this amounts to the identity
\[
R_mT^jR_n=R_{n-j}T^jR_{m+j}, \; \;
\; 1\leq j\leq n+1,
\]
which can be seen as follows
\begin{eqnarray*}
R_mT^j R_n &=& q^{(-1)}\cdots q^{(-m)}q^{(-1+j)}\cdots q^{(-n+j)}\\
   &=&q^{(-1)}\cdots q^{(-m)}q^{(-1+j)}\cdots qq^{(-1)}\cdots q^{(-n+j)}=\\
   &=&q^{(-1)}\cdots q^{(-n+j)}q^{(-1+j)}\cdots qa^{(-1)}\cdots q^{(-m)}\\
   &=&R_{n-j}T^jR_{m+j}.
\end{eqnarray*}
Thus, we conclude that the Miura map is canonical in the sense of the second
Hamiltonian structures.

\section{Conclusions and Discussions}
We have proved that the canonical property of Miura map holds between the lKP
hierarchy and the lMKP hierarchy, that is, it maps the bi-Hamiltonian structures
of the lMKP hierarchy to those of the lKP hierarchy. In \cite{bak2}, the lattice KP
hierarchy is extended and it turns out that the extended lattice KP hierarchy
is isomorphic to the lattice MKP hierarchy. Since we are dealing a slight
generalized version of lMKP hierarchy here (\ref{lmkp}), we  have a different extended lKP
hierarchy.

Introducing a new field $u$ and define the following invertible transformation
\[
u=q,\;\;\; A_0=a_0,\;\;\; A_i=R_ia_i,
\]
It is easy to see that the first Hamiltonian matrix operator for our extended lKP
hierarchy reads 
\begin{equation}
B_{1}^{elKP}=\bordermatrix{ &u&A_0&A_{m>0}\vspace{0.3cm}\cr
               u&0&u(T-1)&\alpha u(T-1)T^mR_m\cr
               A_0&(1-T^{-1})u&0&(T^m-1)A_m\cr
               A_{n>0}&\alpha R_nT^{-n}(1-T^{-1})u&A_n(1-T^{-n})&T^mA_{n+m}-A_{n+m}T^{-n}},
\end{equation}
and the flow equations are given
\begin{eqnarray*}
u_t&=&u(T-1)\frac{\delta H}{\delta A_0}+\alpha\sum_{m=1}^{\infty}u(T-1)T^m\frac{\delta H}{\delta A_m},\\
A_{i,t}&=&\alpha R_nT^{-n}(1-T^{-1})u\frac{\delta H}{\delta
u}+\sum_{j=0}^{\infty}(B_{1}^{elKP})_{ij}\frac{\delta
H}{\delta A_j},\;\; H\equiv H_{n+1}=\frac{1}{n+1}{\rm tr}(L^{n+1}),
\end{eqnarray*}
where the Hamiltonian $H$ is the seam as in the lKP case. We could have a second Hamiltonian
structure for the extended lKP hierarchy, but it is in a rather complicated form. So we omit
it.

To conclude the paper, we point out that it seems interesting to prove the canonical
property of the Miura map on the level of the Poisson tensors since that will hopefully
make the proof more concise. For
the Gelfand-Dickey
hierarchy,
such proof was given by Dickey (\cite{dickey}) and for the continuous KP hierarchy and
the constrained KP hierarchy, it is
provided in 
\cite{shaw} and in \cite{liu1}\cite{st} respectively.

\bigskip

{\bf Acknowledgements}

It is a pleasure to thank  Professor S.Y. Lou for helpful discussions.
The part of work was done during the author's stay in the Abdus Salam
International Centre
for Theoretical Physics and he
should like to thank the AS ICTP for 
hospitality. The work is supported in part by National Natural Science Foundation of
China. 



\begin{thebibliography}{99}

\bibitem{af} M. Antonowicz and A. P. Fordy, Commun. Math. Phys. {\bf 124 }465 (1989).
\bibitem{cheng} Y. Cheng, Commun. Math. Phys. {\bf 171} 661 (1995).

\bibitem{dickey} L. A. Dickey, Soliton Equations and Hamiltonian Systems,
 Singapore: World Scientific (1991).
\bibitem{ds} V.G.  Drinfeld, V.V. Sokolov, J. Soviet Math. {\bf 30} 1975 (1984).

\bibitem{feher} L. Feher, J. Harnad and I, Marshall, Commun. Math. Phys. {\bf 154} 181 (1993). 
\bibitem{bw} B.A. Kupershmidt, G. Wilson, Invent. Math. {\bf 62} 403 (1981).
\bibitem{bak1} B.A. Kupershmidt, Discrete Lax Equations and Differential -
 Difference Calculus, Paris: Asterisque(1985).
\bibitem{bak2} B.A. Kupershmidt, Commun. Math. Phys. {\bf 167}
351-371(1995).
\bibitem{liu} Q. P. Liu, Phys. Lett. A {\bf 187} 373 (1994).
\bibitem{liu1} Q. P. Liu, Inver. Prob. {\bf 11} 205 (1995); Lett. Math. Phys. {\bf 43} 65 (1998).
\bibitem{miura} R. Miura, J. Math. Phys. {\bf 9} 1202 (1968).
\bibitem{oevel} W. Oevel, in Algebraic Aspects of Integrable Systems: in memory
 of Irene Dorfman, eds. A. S. Fokas and I. M. Gelfand, Birkh\H{a}user Boston (1997).
\bibitem{shaw} J.C. Shaw and M.H. Tu, J. Phys. A: Math. Gen. {\bf 30}
4825-33 (1997).
\bibitem{st} J.C. Shaw and M. H. Tu, J. Phys. A: Math. Gen. {\bf 30} L725 (1997).
\bibitem{suris} Y.B. Suris, Phys. Letts. A {\bf 180} 419 (1993).
\end{thebibliography}
\end{document}